\documentclass[conference]{IEEEtran}
\usepackage{cite}
\ifCLASSINFOpdf
\usepackage[pdftex]{graphicx}
\usepackage{amssymb}
\usepackage{lineno}
\usepackage{lscape}
\usepackage{tabularx}
\usepackage[gen]{eurosym}
\usepackage{multirow}
\usepackage{hhline}
\usepackage{booktabs}
\usepackage{siunitx}
\usepackage{float}
\usepackage{framed}
\else
\fi

\begin{document}
%

\title{On the Relation Between Unit Testing and \\Code Quality}




\author{\IEEEauthorblockN{Lucas Gren and Vard Antinyan}
\IEEEauthorblockA{Chalmers University of Technology and the University of Gothenburg\\
Gothenburg, Sweden 412--92 \\
Email: lucas.gren@gu.se and vard.antinyan@gu.se}
}

\maketitle

\begin{abstract}
Unit testing has been considered as having a key role in building high quality software, and therefore it has been widely used in practice. However, data on the relationship between unit testing and aspects of software quality remain scarce. A survey study with 235 survey responses from seven organizations was conducted in order to understand the correlation between practitioners' perception of code quality and unit testing practices. In addition, we conducted a case study in one of these organizations to investigate the correlation between unit test coverage and post-unit test defects. In both cases none or weak correlations were found. We recommend further research on the effectiveness of different testing practices in order to help practitioners to understand how to best allocate their resources to the testing chain.
\end{abstract}


\IEEEpeerreviewmaketitle

\section{Introduction}\label{sec:introduction}
Unit testing has gotten a key place in modern software development. In a survey by Runeson \cite{runeson2006survey} the results show that the main reason for starting to use unit testing was external requirements. The test suites could then function as a technical specification. Another common reason for using unit tests was to proclaim the use agile methods in the organization. Several practitioners have expressed issues with not having measures to validate the usefulness of unit testing. This hinders estimating a return on investment (ROI) in unit testing and motivating developers to run thorough unit tests \cite{blogpost2}. Currently, unit testing is promoted in many international standards such as ISO 26262, which strongly recommend thorough unit testing in safety critical systems. Companies that develop safety critical software, such as automotive or air traffic control systems, need to adhere to these standards, because using them provides confidence for developers and gives credibility in the eyes of the customers. Other companies, that do not have safety critical software, are more free in choosing how much unit testing that should be conducted, which means it is up to individual developers or software development teams to decide. Empirical data on how unit testing influences the quality for either types of software do not exist, and therefore, practitioners are not able to make evidence-based decisions.

We know that the usefulness of unit testing is a debated issue in the software development community with strong opinions for their extreme usefulness, on one hand, and their complete uselessness on the other. To us, such a debate indicates a lack of empirical data on the subject, and therefore, this study tries to provide data on some aspects of the the relationship between unit testing and software quality, namely that of test coverage and number of defects, and perceived code quality and unit testing practices. Therefore, the research questions addressed in the study are as follows:

\begin{enumerate}
\item To what extent is unit test (statement\slash branch) coverage correlated to number of defects?
\item To what extent is the degree of unit testing correlated to perceived code quality?
\end{enumerate}



\section{Background}\label{background}
Unit testing has been believed to be one of the pillars of code quality over a long period of time. There are thousands of online blogs and discussion forums created by practitioners, with statements that unit testing is one of the success factors in developing high quality code. Researchers, in their turn, have not pulled their punches in proposing different techniques for unit testing in practice \cite{hamill2004unit}. The reasons for this is probably the \emph{a priori} knowledge in software engineering that asserts the necessity of unit testing in software development; at university, we are taught that unit tests are the cornerstones of code quality, and in industry, we feel safer if the unit test coverage is high. One of the highly read and positively rated posts (with over 700 votes) summarizes the online discussions well: \emph{Question}: ``What are some good ways to convince the skeptical developers on the team of the value of unit testing?'' \emph{Answer}: ``I just love unit tests, I have just been able to make a bunch of changes to the way something works, and then was able to confirm I had not broken anything by running the test over it again...'' \cite{blogpost1}. This answer encapsulates the following arguments that are widely stated in such discussions; unit testing helps understanding the design and code changes, it gives visual feedback, helps with documenting and reusing code, and does not slow down the coding, as it might be perceived to do, at first glance. 

There is also a relatively small community of developers that believe otherwise. One of the top developers at one of the participating companies said: ``I am afraid some of our developers conduct unit testing, not because it ensures high product quality, but because it gives the feeling of that they did all that was needed to ensure high product quality.'' According to a researcher at MIT one of the top developers of Google was even more strict in his judgment, stating that: ``the most help he could ever get from software engineering research would be if somebody could show something he had been sure of for ten years; that unit testing does not give anything, all defects are found in integration and system testing anyways'' \cite{googlequote}.

There is evidence showing that a Test-Driven Development (or TDD) approach finds significantly more defects than a Test-After approach \cite{maximilien2003assessing}, but it is unclear if this effect arises from better code comprehension through TDD or from the unit testing itself. In fact, since the TDD approach has such a large effect, as compared to regular unit testing, this might be an indication that there is low value in unit testing itself. One study using number of passed acceptance tests as a measurement of external quality showed no increase in quality by writing tests first \cite{zielinski2005preliminary}.

Despite that many are positively inclined towards the value of unit testing, and others believe the complete opposite, scientific evidence on this matter is surprisingly scarce. Few existing studies indicate that unit testing might not have a significant connection to quality. One of the studies conducted by Mockus et al.\ \cite{mockus2009test} shows that the correlation between unit test coverage and defects is none or very weak. In that study, test coverage was used as a measure of unit testing sufficiency and defect count was used as a measure of quality. In fact, the authors did not focus on quality and unit testing, but explicitly stated that they investigate the relationship between known pre-unit test defects and unit test coverage. However, considering their other finding, that is, a linear increase in coverage requires an exponentially increase in effort, we can deduce that an increased effort in unit testing did not pay off with regards to decreasing defects. We find this result remarkable, because the research setup was in a context of real product development, with a minimal set of manipulated conditions. They had two large industrial products, actual defects, and controlled confounding factors in their data analysis, such as code size and changes. There are other studies, which also indirectly relate to unit testing and defects correlation (see for example \cite{gay2015risks}). However, based on these studies it is difficult to assess the relationship between unit tests and quality, because they are not sufficiently close to a real case. Particularly, the following issues pertain: (1) Artificial defects (mutants). (2) Uncontrolled confounding factors such as size and change rate. (3) Artificially manipulated range of unit test coverage. (4) Small products and a low number of defects. Alongside these quantitative studies there are also few qualitative studies on the relationship between unit testing and product quality. For example, contradicting responses were obtained from interviews of 605 software engineers, indicating that even though the majority of respondents use automated unit testing because of various reasons or beliefs, 52\% of the respondents did not consider unit testing as a test sufficiency criterion \cite{lyu1994coverage}.

The study is composed of two independent parts; the first part being a survey conducted in industry with two companies in Sweden, and three companies and a university in Brazil, and the second part being a case study in a large company in Sweden.

\section{Method}\label{Method}

\subsection{Survey context}
The sample consisted of data from both Brazilian and Swedish companies, comprising IT departments at one large online media and social networking company with around 5,000 employees, one smaller software life-cycle consultancy company with around 35 employees, and one company that offers programming courses to individuals and companies (around 100 employees), one multinational networking and telecommunications equipment and services company (with around 115,000 employees), an aerospace and defence company (with around 14,000 employees), and an automotive parts manufacturing company (with around 160,000 employees). We also collected data from the University of S\~ao Paulo, consisting of software engineering students enrolled in an agile software development (XP) project course. The surveys were given to 241 software development team members and 201 responded, which gives the total response rate of 83\%. We also collected data a second time after two months trying to match individual responses, and we managed to pair 49 software development team member responses in a repeated measurement. 


\subsection{Case study context}
The case study was conducted at the multinational networking and telecommunications equipment company, mentioned in the previous section. The product contained over two million lines of code, developed by distributed semi-autonomous development teams. The number of software engineers in the organization was about 150. The development was conducted based on agile principles and continuous integration methods. The organization employed automated unit testing before code was delivered to the main code branch of the product. The organization used statement and branch coverage of unit tests as a guide for test sufficiency, however, there was no strict decision criteria defined for coverage. The design architects and development teams decided themselves how much unit testing should be conducted, and therefore, various amounts of unit tests were written in different parts of the product.

\subsection{Survey data collection and analysis}
To measure software development team members' perception of the degree of unit testing we used a survey developed by So and Scholl \cite{so}. It is the only survey we found that is validated through a factor and reliability analysis ($N=227$). However, in this study we only used three items of that questionnaire, namely the ones regarding unit testing: (1) The implemented code was written to pass the test case. (2) New code was written with unit tests covering its main functionality. (3) All unit tests were run and passed when a task was finished and before checking in and integrating.

All team members were given the possibility to answer the survey no matter their role, however, they were explicitly told to skip questions the did not have good knowledge about. The respondents were asked to provide responses for each of the items on a Likert scale ranging from 1 (never) to 7 (always). Perceived code quality was measured by the single question: ``How would you rate the code quality in your product(s)?''

The quality question was evaluated on a Likert scale from 1 (Very Poor) to 5 (Excellent). The three items on unit testing and the fourth item on quality were not the only items in the provided survey. The whole questionnaire was much longer, containing all items from the Perceptive Agile Measurement as suggested by \cite{so}, therefore, there was a low probability that respondents could make an explicit connection between the items on unit testing and quality, because they responded to the unit testing items on average ten minutes before assessing the quality item. The survey was distributed in paper form and collected on site on two occasions, hence the high response rate.

To evaluate if the data was normally distributed, we plotted frequency diagrams for all the four factors. There were non-normal patterns in the frequency plots, and, therefore, we ran the Shapiro-Wilk test for normality. The test statistics were significant for all factors (Quality item: Test statistic = 0.845, $p$ = 0.000, Unit Tests 1: Test statistic = 0.932, $p$ = 0.000, Unit Tests 2: Test statistic = 0.905, $p$ = 0.000, and Unit Tests 3: Test statistic = 0.871, $p$ = 0.000), i.e.\ we had an issue with the normality assumption. Since we also have values that have the same ranking, the Spearman rank correlation coefficient was used as a measurement of correlation, because it allows tied ranks in data. Because we did not know the direction of the correlations we opted to use two-tailed tests. We also tested if there was a difference between students ($N=35$) and practitioners ($N=151$). The only significant Mann-Whitney U test was using the third unit testing item, namely: ``All unit tests were run and passed when a task was finished and before checking in and integrating.'' with $U= 1,941.5, p=0.006$, i.e.\ students rated that they do less of running and passing test cases before checking in and integrating. A possible explanation for this could be the fact that they build new products from scratch and therefore were less thorough with unit tests before each check-in or integration to the main branch.







In order to try to investigate causality further, we conducted a second data collection at the Brazilian companies two months after the first measurement. However, having the participants provide a personal identifier which maintained their anonymity proved to be difficult. After having collected the second measurement we managed to pair 49 individual responses and ran a related-samples Wilcoxon signed rank test.

\subsection{Case study data collection and analysis}
In order to being able to investigate the relationship between unit tests and code quality we used unit test coverage and post-unit-test defects in files respectively. Statement and branch coverage were used as a measure of unit test coverage \cite{myers2011art}. Statement coverage is the percentage of source code statements that has been exercised during a test run. Branch coverage is the percentage of decision blocks in a file that has been exercised during a test run. 

The defects per file were measured by the number of modifications that was tagged as a bug fix in the version control system. We believe that this method for measuring defects was accurate, because the regulations in the organization obliged developers to tag all bug fixes in the version control system. The analysis is conducted under the assumption that if unit tests are effective in finding defects, then high unit test coverage should provide a low rate of post-unit-test defects. The correlation coefficient between statement\slash branch coverage and defects is therefore expected to be negative. Moreover, a higher value of the squared correlation coefficient can be seen as a measurement of effect size and indicates how much variance in defects that can be explained by unit test coverage. 

As an additional cautionary step, we chose to only use files that have either 100\% or 0\% coverage. The reason was that if we use files with other coverage values as well, the size of the files would intervene with the accuracy of the analysis. Such an analysis would then assume that, if unit tests are effective, then files with equal amount of coverage should have nearly equal amount of defects. However, this assumption is not true because a file that has 1000 lines of code and 50\% coverage is likely to have more defects than a file with 100 lines of code and 50\% of coverage. This is because 1000 lines of code with 50\% coverage has 500 lines of untested code, which is much higher than the second file's 50 lines of untested code. For this reason a reasonably simple analysis of defect-coverage relationship cannot be done. For files that have anything but 100\% or 0\% coverage a normalization of coverage over size (complexity and changes) would be required. This kind of normalization, however, is not as straightforward as it may seem. For that reason we left such analysis for an upcoming study of ours where data from multiple products will be collected and it would be possible to present data from different angles.

Both the number of defects and the statement\slash branch coverage were measured for one major release of the product, which was about one year in duration. The coverage was measured at two points in time (two snapshots of the code in the main integration branch), the first time was the very beginning of the development of a newly planned release of the product and the second time was at the end when the product was ready for release. We did the two measurements in order to check whether the test coverage for files changed over time. By comparing the coverage per file at two occasions, we observed that the coverage for the majority of files (97\% of the cases) was constant over time. This meant that we could conduct a meaningful analysis of coverage-defect relationship for the 97\% of the files. The few files that had varying coverage were excluded from the analysis.

We also measured the number of post-unit-test defects for the same period. These defects included all the defects that came from unit tests and were found in integration tests, system tests, or were reported by customers when using the small test-releases of the product. Thus we had all the files that had constant statement and branch coverage (either 0\% or 100\%) over one year of development, and we had the number of post-unit-test defects reported per file.

We decided to use correlation analysis in order to assess the correlation between coverage and number of defects. The defect count is a continuous variable, because a file can have several defects, and it has an absolute zero meaning ``no defects.'' The coverage is a dichotomous variable, because it can have only two values -- 0\% and 100\%. To visually understand the correlation between the defects and coverage, and to evaluate normality of the defect data we used a marginal plot of the two variables. The plot revealed concerns with the normality assumption, therefore a Shapiro-Wilk test of normality was conducted (Test statistic = 0.435, $p$ = 0.000). As the results show, the test was significant, i.e.\ we have an issue with normality on this variable too. Thus, we need to fulfill the following conditions when choosing correlation analysis technique: (1) One of the variable is continuous and the other variable is dichotomous. (2) The continuous variable is not normally distributed.


For these conditions the most appropriate correlation analysis technique is rank bi-serial correlation, which we applied to assess the correlation between the number of defects and statement\slash branch coverage.

The last concern was the possibility of confounding factors that could affect the results of the correlation analysis. The first concern was whether there is a practitioner's bias when choosing which files to test and which files not to test. If the practitioners consciously or subconsciously make a choice to test files that are more complex (bigger, simpler, etc.) then the results would be affected and misleading. For these reason we organized three meetings with 10-14 software designers in order to understand how they choose the extent of  testing. Before the meetings the software designers were not informed of this current study. The summary of the discussions showed that the testing largely depends on individual software developers' own conviction of how much unit tests that should be written. In some cases the design architects of certain code areas could also recommend certain amount of testing. We did, however, not find any indication that the software designers' choices were dependent on factors like complexity, size, or perception of error-proneness. This particular finding is congruent with findings of Daka et al. \cite{daka2014survey}.

The second concern is how much size, complexity, evolution, and other factors affect the correlation analysis results directly. Since we have a fairly large number of files (680), and since the number files with 0\% coverage is comparable with the number of files with 100\% coverage (see histogram in Figure~\ref{mp}), we assume that the effect of these confounding factors on the measured variables is random. Simply stating, we assume that there is an equal amount of simple and complex files in both the 0\% and the 100\% groups. These facts reduce the risk that the confounding factors have a large impact on the analysis results.

\section{Results}\label{sec:results}
The correlations between the four survey items can be seen in Table~\ref{fig:corrmatrixapa}. The correlation between the Quality item and Unit Tests 1 was \emph{not} significant. The two other unit tests items were significantly correlated with the coefficients $0.321$ and $0.226$ respectively, at an alpha level of less than 1\%. If we square these correlation coefficients, we can get a measure of explained variance in the correlation model \cite{coheneffect}. In this case, we then have effect sizes of 10\% and 5\% percent.

\begin{table}
\renewcommand{\arraystretch}{2}
\caption{Spearman's $\rho$ Correlations. Perceived Code Quality and Unit Tests ($N=186$)}
\label{fig:corrmatrixapa}
\centering
\begin{tabular}{cccccc||c||c||c||c||c||c||c}
\hline
Measure \bfseries & 1 \bfseries & 2 \bfseries & 3 \bfseries  &4  \\
\hline
1. Rate the code quality in your product(s) & 1 & 0.169 & 0.321* & 0.226*\\
\hline
2. Implemented code written to pass test case &  & 1 & 0.507* & 0.405*\\
\hline
3. Main functionality unit tests coverage &  &  & 1 & 0.596*\\
\hline
4. Unit tests passed when before integrating &  &  &  & 1\\
\hline
{*p$<$.01 (2-tailed)}  \\
\end{tabular}
\end{table}

We did not find any significant difference between the first and the second measurement, even if we managed to block the effect of individuals with a related-samples Wilcoxon signed rank test for all the three unit testing items (Unit Tests 1, Test Statistic $= 428, p=0.125$, Unit Tests 2, Test Statistic $= 351, p=0.099$, Unit Tests 3, Test Statistic $= 334.5, p=0.326$) and the quality question (Quality item, Test Statistic $= 105, p=0.142$). It is hard to draw any other conclusion than that it apparently takes longer time for these items to change significantly in a real development context. 

\begin{figure}
\centerline{\includegraphics[scale=0.42]{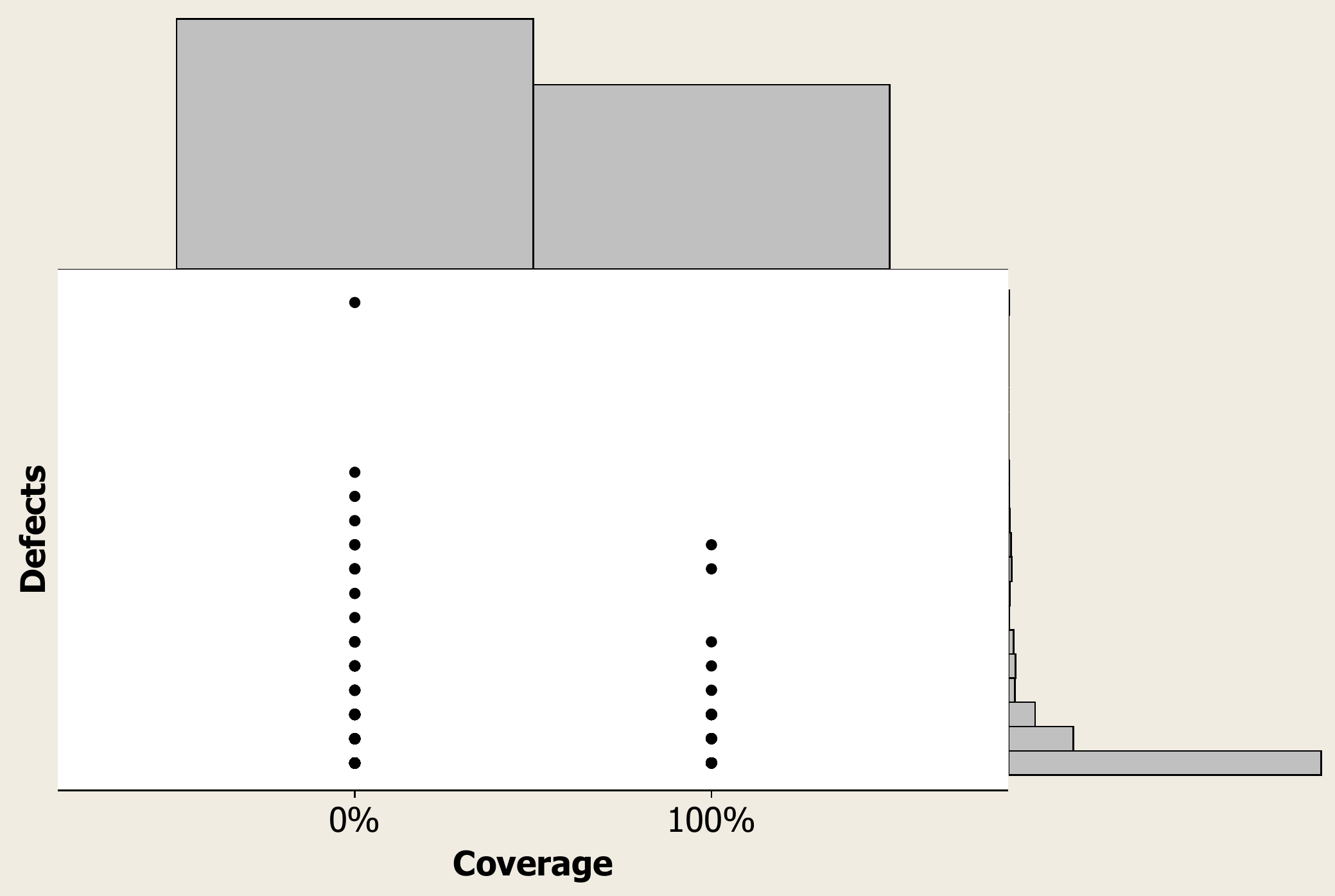}}
\caption{Marginal plot for defects and coverage.}
\label{mp}
\end{figure}

Figure~\ref{mp} shows a marginal plot of defects and unit test coverage. The histogram on the right-hand side of the figure shows that the distribution of defect does not look normal. This makes sense since most of the files are not expected to have defects. The scatter plot shows the relation of defects and coverage. By looking at the plot we can see that both groups of files contain defects, and moreover, several files with 100\% coverage have multiple defects. Having so many data points (680) usually gives a significant results of statistical tests, but then the effect size is what needs to be analyzed. In fact, the correlation analysis results in Table~\ref{DC} show that the effect size of the correlation between defects and statement\slash branch coverage is very small (2.9\% and 3.6\%).

\begin{table}
\renewcommand{\arraystretch}{2}
\caption{Rank biserial correlation ($N=680$).}
\label{DC}
\centering
\begin{tabular}{ccccc||c||c||c||c||c||c}
\hline
Correlation of \bfseries & Cor. coefficient \bfseries & Effect size \bfseries & p-value  \\
\hline
Defects vs. statement coverage & -0.17 & 2.9\% & 0.000* \\
\hline
Defects vs. branch coverage & -0.19 & 3.6\% & 0.000* \\
\hline
{*p$<$.01 (2-tailed)}  \\
\end{tabular}
\end{table}

\section{Discussion}\label{sec:discussion}
Regarding the first research question, we found no correlation between the first item of unit testing and perceived code quality, and we found only low correlations between the second and third unit testing items and perceived code quality. This indicates that if unit testing is poorly practiced in an organization, the product quality does not necessarily have to be low. In fact, the correlation between the first item of unit testing practices and perceived code quality was not statistically significant, and the effect sizes between the other two unit testing practices and code quality were 10\% and 5\% correspondingly. Because there were small or no correlations between perceived code quality and unit testing practices, these results do not support any causation between these two constructs (since causation is unlikely without correlation). 

Furthermore, regarding the second research question, the measurements in the case study gave low correlation between defects and coverage. This result shows, that no matter if files have 100\% or 0\% unit test coverage, the differences in defects between these two groups only showed a small effect. The effect size was as low as 2.9\%, therefore, the assumed causal relationship between unit tests and code quality is dubious.

The effect size of 2.9\% means that 2.9\% of the variation in defects could be explained by the unit test coverage. Knowing that companies might spend up to 40\% of their resources on coding and unit testing \cite{yang2008phase}, one would expect an effect size of considerably higher magnitude, in order to justify the spent time on unit testing, however, this of course depend on the system being built.

Our current study is only a first step, and more empirical studies must be conducted in order to pin down the usefulness of unit testing in practice. Particularly, the following issues should be meticulously investigated: 1) the relation between unit testing and code quality should be investigated in different domains and development methodologies, 2) the quality of the unit tests can vary and therefore sheer correlation between the coverage and defects can underestimate the effect size 3) it is maybe so that coverage measures are inadequate measures for measuring unit testing, 4) code complexity and cohesion have major effect on code quality \cite{buse2010learning}, therefore they are likely to be strong confounding factors, 5) the developers choice of which files should be tested can be deliberate.


\subsection{Threats to Validity}
In case of the survey both measures are derived from the software engineers' perception of code quality and degree of unit testing. Therefore the real connection between quality and testing could be different \cite{devanbu2016belief}: The summary of experienced engineers' perception is useful but limited.

In the case study, statement and branch coverage measures were used as measures of unit testing sufficiency. Statement coverage and branch coverage have been criticized in the literature for being inaccurate measures of test sufficiency \cite{marick1999misuse}, and therefore the correlation between coverage and defects might be significantly different from the actual correlation of unit testing and defects. We mitigated this threat by using only files in the two extremes, i.e.\ files that had either 0\% or 100\% unit test coverage. In this case the threat that coverage measures would be inaccurate is then less of a concern, because the difference in effort between 0\% and 100\% is substantial. Hence, if 100\% coverage cannot help with reducing post-unit test defects significantly, then why to write unit tests at all.

In this paper we only have one case study and one survey, while unit testing practices across different domains and programming languages can be considerably different. Particularly, in safety critical systems unit testing is conducted more rigorously. There is a pivotal difference between unit testing of hand-written and generated code (e.g.\ generated from Simulink or Rhapsody models). In the case of generated code, unit testing is quite similar to black box testing, because testing is done on models where only input and output signals are relevant. 

\section{Conclusions and future work}\label{sec:future}
This study set out to investigate if unit test coverage is correlated to number of defect and if perceived code quality is correlated to unit testing practices. Through a survey and a case study we found data that do not support a strong causal relationship between unit testing and code quality, since we did not find strong correlations. The results of this study suggest that the effect of unit testing on code quality is questionable, therefore, more studies with industry data need to be conducted in order to get conclusive results on the effect of unit testing on code quality.


\bibliographystyle{IEEEtran}
\bibliography{references}
\end{document}